\documentstyle[twoside,12pt] {article}                                                   
\begin{document}                                                                
%
%Definitions:
\def\ie{{\it i.e. \/}}
\def\eg{{\it e.g. \/}}
\def\cf{{\it c.f \/}}
\def\viz{{\it viz.\/}}
\def\cap{\caption}
\def\cline{\centerline}
\def\half{{\scriptstyle{\frac{1}{2}} }}
\def\halft{\textstyle{\frac{1}{2}}} 
\def\osqrt{\textstyle{\frac{1}{\sqrt2}}} 
\def\lsqrt{\textstyle{\frac{\l}{\sqrt2}}} 
\def\phalf{\textstyle{\frac{\pi}{2}}}  
\def\nd{\noindent}
\def\nn{\nonumber}
\def\ad{a^\dagger}
\def\ab{\bar{\alpha}}
\def\nub{\bar{\nu}}
\def\hi{\chi_{klm}}
\def\udp{U_{\lambda}^{\dagger}}
\def\udm{U_{-\lambda}^{\dagger}}
\def\utp{\tilde{U}_{\lambda}}
\def\utm{\tilde{U}_{-\lambda}}
\def\up{U_{\lambda}}
\def\um{U_{-\lambda}}
\def\so{\sigma_{1}}
\def\st{\sigma_{2}}
\def\sth{\sigma^{z}}
\def\sp{\sigma^{+}}
\def\sm{\sigma^{-}}
\newcommand{\fscr}[2]{\scriptstyle \frac{#1}{#2}} 
\newcommand{\bra}[1]{\left<#1\right|}
\newcommand{\ket}[1]{\left|#1\right>} 
\newcommand{\braket}[1]{\left<#1\right>}
\newcommand{\inner}[2]{\left<#1|#2\right>}
\newcommand{\sand}[3]{\left<#1|#2|#3\right>}
\newcommand{\proj}[2]{\left|#1\left>\right<#2\right|}
\newcommand{\absqr}[1]{{\left|#1\right|}^2}
\newcommand{\abs}[1]{\left|#1\right|}
\newcommand{\lag}[2]{L_{#1}^{#2}(4\l^{2})}
\newcommand{\dbl}[2]{\rm#1\hskip-.5em \rm#2}
\newcommand{\mes}[1]{d\mu(#1)}
\newcommand{\pl}[1]{\partial_{#1}}
\newcommand{\mat}[4]{\left(\begin{array}{cc}
                           #1 & #2 \\
                           #3 & #4
                          \end{array}\right)}
\newcommand{\col}[2]{\left( \begin{array}{c}
                                 #1 \\
                                 #2
                            \end{array} \right)}
\def\a{\alpha}
\def\b{\beta}
\def\g{\gamma}
\def\d{\delta}
\def\e{\epsilon}
\def\z{\zeta}
\def\th{\theta}
\def\f{\phi}
\def\l{\lambda}
\def\m{\mu}
\def\p{\pi}
\def\om{\omega}
\def\D{\Delta}
\def\zb{\bar{z}}
\newcommand{\be}{\begin{equation}}                                              
\newcommand{\ee}{\end{equation}}                                                
\newcommand{\ba}{\begin{array}}  
\newcommand{\ea}{\end{array}}                                                
\newcommand{\bea}{\begin{eqnarray}}                                              
\newcommand{\eea}{\end{eqnarray}}                                                
\newcommand{\beann}{\begin{eqnarray*}}                                              
\newcommand{\eeann}{\end{eqnarray*}}                                                
\newcommand{\bfg}{\begin{figure}}                                                
\newcommand{\efg}{\end{figure}}                                                
\begin{titlepage}                                                               
%\hspace*{10.0cm}\parbox[t]{3.0cm}{FTUV/94-42\\IFIC/94-47\\              
%September 1994}
\begin{center}
Published in Physica D, {\bf 134}, 126 - 143 (1999)
\end{center}
\vskip 1.0 cm                                                                   
\begin{center}                                                                  
{\Large\bf Quantum nonlinear lattices and coherent state vectors} 
\vskip 1.0 cm                                                                   
                                                                                
{\bf Demosthenes Ellinas $^1$}\renewcommand{\thefootnote}{*}\footnote{     
Email: {\tt ellinas@science.tuc.gr}} ,
{\bf Magnus Johansson $^2$}\renewcommand{\thefootnote}{\diamond}\footnote{     
Email: {\tt mj@imm.dtu.dk}}
and \
{\bf Peter L Christiansen $^2$}\renewcommand{\thefootnote}{\natural}
\footnote{     
Email: {\tt plc@imm.dtu.dk}}
\vskip 0.2cm
$^1$ Department of Sciences\\
Technical University of Crete \\ 
GR-73 100 Chania Crete Greece\\              
\vskip 0.2cm
$^2$ 
Department of Mathematical Modelling\\
The Technical University of Denmark\\
DK-2800 Lyngby, Denmark
\vskip 2.0 cm                                                                   
{\bf Abstract}                                                                  
\end{center}                                                                    
\vskip 0.2 cm              
\par Quantized nonlinear lattice models are considered for 
two different classes, boson and fermionic ones. The quantum 
discrete nonlinear Schr\"odinger model (DNLS) is our main objective, 
but its so called modified
discrete nonlinear (MDNLS) version is also included, together with 
the fermionic polaron (FP) model. 

\indent  Based on the respective dynamical symmetries
of the models, a method is put forward which by use of the associated boson
and spin coherent state vectors (CSV) and a factorization ansatz for the 
solution 
of the Schr\"odinger equation, leads to quasiclassical Hamiltonian equations
of motion for the CSV parameters. The so obtained 
evolution equations are intimately related to the respective evolution
equations for the classical lattices,
provided we account for the ordering rules (normal, symmetric)
adopted for their quantization.

\indent Analysing the geometrical content of the factorization 
ansatz made for the state vectors invokes the study of the 
Riemannian and
symplectic geometry of the 
CSV manifolds as generalized phase spaces. 
Next, we investigate analytically and numerically the behavior of mean values 
and 
uncertainties of 
some physically interesting observables as well as the modifications 
in the quantum regime of processes such as the discrete self trapping (DST),
 in terms of the Q-function and the 
distribution of excitation quanta of the lattice sites. Quantum DST in the 
symmetric ordering of lattice operators is found to be relatively enhanced 
with respect to the classical DST. Finally, the meaning of the factorization 
ansatz for the lattice wave function is explained in terms of disregarded 
quantum correlations, and as a quantitative 
figure of merit
for that ansatz a correlation index is introduced. This index is given in terms 
of the norm of the difference between the
true and factorized state vectors, and accounts for the quantum correlations
of the lattice sites that develop during the time evolution of the systems. 

\vskip 3.0 cm
\end{titlepage}                                                                 
\section{Introduction}
Nonlinear lattice equations are abound in physical applications and 
their study constitutes a well developed field of research. In many cases where
models are described by classical lattice differential equations this is done
as a first approximation to what is essentially a quantum mechanical lattice 
system. Quantized nonlinear lattice equations are the main target of the most 
successful method of the quantum inverse scattering method (QISM)~\cite{kbi}. 
Recent
attempts to treat quantum lattice equations with arbitrary number of 
lattice sites
 have 
led to such schemes as
the so called number state method (NSM)~\cite{se,esse}, and the Hartree 
approximation~\cite{wehms} (see also~\cite{bbp} for a discussion about 
QISM and NSM).
On the other hand it seems highly desirable when treating quantum versions 
of classical lattices to be able to utilize in some sense 
the available classical solutions of the nonlinear evolution equations and
to keep a conceptual framework \viz phase spaces, Poisson brackets, coordinate
transformations etc. akin to that of the classical models. It is in that
direction that this work focuses and provides an alternative to the 
above mentioned methods.

Specifically we deal with lattices related to the nonintegrable 
generalized discrete nonlinear
Schr\"odinger equation (DNLS)~\cite{els} (also known as the generalized discrete
self-trapping (GDST) equation)~\cite{cs} and with a modified version of it
(MDNLS)~\cite{kto}. We also
discuss a fermionic lattice associated to the fermionic polaron (FP) 
model~\cite{fed}.
The DNLS models quantized by the canonical quantization method lead
to a  configuration of $f$ sites, described by tensoring the 
Weyl-Heisenberg (WH) algebra attached to each lattice site. The ordering rule
adopted when the classical wave amplitude variables are substituted by boson
operators~\cite{dir,weyl} is essential for the form of the 
obtained Hamiltonian operator.
Our method of solving the resulting Schr\"odinger equation is based on the 
assumption that the lattice state vector can be factorized 
to a tensor product of coherent state vectors~(\cite{sch,gla,ber,per,rmp})
 each one living in the 
Hilbert space associated to each lattice site. This leads to a set of 
Hamiltonian equations for the CSV coordinates, which provide an
approximate quasiclassical solution for the exact quantum dynamics. As
the CS are the most 
classical state vectors~\cite{sch,gla} for boson systems the resulting
equations are quasiclassical in nature and govern the motion of CS wavepackets.
To elucidate the meaning of the ansatz we recall that the boson CS labels
are coordinates of the canonical phase plane and therefore the quasiclassical 
dynamics occurs in the $f$-fold cartesian product of harmonic oscillator
phase space. That dynamics is the classical dynamics if the normal ordering
of the bosons is used and diverges from that for symmetric ordering. Therefore 
the CS mean values for normally ordered observables coincide with their 
classical values in the former case while their quantum mechanical fluctuations
are constant and minimum in the course of time evolution.
To try the method to more general quantum systems we have also studied the
FP model, which is first mapped by means of the Jordan-Wigner transformation
to its equivalent XXZ model~\cite{xxz}. The resulting spin Hamiltonian is 
expressed by
tensoring the $su(2)$ algebra generators attached to each site of the
model. By the same token as in the boson case the quantum dynamics is
studied by assuming factorization  of the state vector of the system to a 
product of spin
CS. However the geometrical framework for the dynamics of the spin CSV's
is different in this case. The $su(2)$ CS space is 
a spherical surface with a noncanonical symplectic structure defined on it,
as it results by looking at the so called ray metric of the 
CSV~\cite{kib,pro}; the classical
Hamiltonian equations generate a flow in the $f$-fold cartesian product of 
spherical (or rather complex projective) coordinates.

The plan of the paper is as follows: in section two the general framework
for the construction of CSV is outlined and the needed formulas
for the special cases of the canonical and the spin CSV are provided.
In the next section the DNLS
model is quantized and the associated quasiclassical equations
are derived and studied. Attention is paid to the problem of discrete self
trapping (DST) phenomenon in the quantum regime, by illustrating the situation
with plots of the Poisson distribution and the Q-function of the
quantized modes of the system. Next section explains the physical meaning of
the factoring of the lattice state vector as adopted for all models, and gives
a validity measure for that approximation. In the following two appendices
we illustrate the same idea by taking up first the MDNLS quantum lattice and
then the FP model.
The final section contains a number of conclusions and 
offers some prospects.

Throughout the text we consider $\hbar=1$ and we denote CS 
expectation values of any generator with brackets around it.
\section{Coherent state manifolds}
The notion of coherent state vector was essentially introduced in the 
early days of quantum mechanics by Schr\"odinger~\cite{sch}, in his
succesful attempt to construct a nonspreading Gaussian wavepacket
for a quantum harmonic potential. The wavepacket center was evolving
following the classical path while the dispersion was kept to a
minimum compatible with the minimum uncertainty principle. 
The regeneration of the theory took place in the sixties and 
seventies when to a wealth of physical applications of CS
a mathematical group theoretical foundation was 
also provided~\cite{kla,gil,per,rmp}. 

\noindent For our needs here a brief introduction of the CS concept
goes as follows: consider a Lie group $\cal G$, with a unitary irreducible 
representation $T(g)$, $g\in \cal G$, in a Hilbert space $\cal H$. We
select a reference vector $\ket{\Psi_0}\in \cal H$, to be called the 
"vacuum" state vector, and let ${\cal G}_0\subset \cal G$ be its isotropy 
subgroup, \ie for $h\in {\cal G}_0$, $T(h)\ket{\Psi_0}=
e^{i\varphi(h)}\ket{\Psi_0}$. The map from the factor group 
${\cal M} = {\cal G}/{\cal G}_{0}$  to the Hilbert space 
$\cal H$, introduced in 
the form of an orbit of the vacuum state under a factor group element,
defines a CSV 
$\ket{x}=T({\cal G}/{\cal G}_{0})\ket{\Psi_0}$ labelled 
by points $x\in\cal M$ of the coherent state manifold.
Coherent states form an (over)complete set of states, since 
by means of the Haar invariant measure of the group $\cal G$ \viz
$\mes{x}, \;\; x\in \cal M$, they provide a resolution of unity,
$\bf{1}=\it{\int_{\cal M}\mes{x}\proj{x}{x}}$. As a consequence, any vector
$\ket{\Psi}\in\cal H$ is analyzed in the CS basis, 
$\ket{\Psi}=\int_{\cal M}\mes{x}\Psi(x)\ket{x}$, with coefficients 
$\Psi(x)=\inner{x}{\Psi}$.

\noindent What concerns us here mostly is the geometry of the CS manifold
$\cal M$. Indeed by its very construction $\cal M$ inherits the structure of 
a Riemann manifold with in general non-constant 
curvature, which is also 
endowed with a complex structure of a
K\"ahler manifold~\cite{h}, namely it can be considered as a 
generalized phase space~\cite{ber}.
In the sequel we restrict ourselves to the case of a two dimensonal surface
$\cal M$ for definiteness, although higher dimensional extensions of our 
statements are also possible~\cite{per}. Also we shall assume a generator
$\rm G_+$ creating from the vacuum state the CS vector, {\it i.e.},
\be
\ket{\z}={\cal N}||\z)={\cal N}{\rm exp}(\z {\rm G_+})\ket{\oslash}\;,
\ee
\noindent where $\z\in C$ and 
${\cal N}=(\z|| ||\z)^{-\frac{1}{2}}$ 
is the normalization factor. 
Occasionally we shall write $\rm G_-$ for 
the Hermitean conjugate of $\rm G_+$; they should both
belong to the Lie algebra of the group for which the CS is defined. 
Below we shall
specialize to the cases of $su(2)$ and the Weyl-Heisenberg groups 
but for the moment we
proceed with the present general framework.

\noindent The symplectic structure possessed by the state space $\cal M$ is 
based on the existence of a canonical kinematical 1-form 
$\theta=\bra{\z}d\ket{\z}=
{\fscr{1}{2}}(\braket{G_+}d\z-\braket{G_-}d\overline\z)$.
The derivation $d=\partial_\z d\z+\partial_{\overline{\z}}d\overline\z$
acts on the state vectors, {\it e.g.}, 
$d\ket{\z}=(G_+ -{\fscr{1}{2}} \braket{G_+})\ket{\z}d\z+
(-{\fscr{1}{2}}\braket{G_-})\ket{\z}d\overline\z $.
\noindent The symplectic 2-form $\om$ of $\cal M$ is derived from the 
canonical $\th$ by derivation $\om=d\th$, and can be expressed in the form 
$\om=(\braket{G_- G_+}-\braket{G_-}\braket{G_+})d\z \wedge  d\overline\z$.

\noindent 
Concerning the geometric features of our $2D$ phase space considered 
as a Riemannian     
surface with a distance function operating on it, we shall employ a 
meaningful metric tensor starting from the       
distance on the projective Hilbert space of rays 
$\overline{\ket{\z}}$, $\z \in  C$~\cite{kib,pro}.
%$\overline{\ket{\z}},\; \z \in \bf{C} $~\cite{kib}-\cite{ana}.      
By $\overline{\ket{\z}}$ we simply mean the set of      
CS $e^{i\phi}\ket{\z}$,                                                      
multiplied by an arbitrary phase factor. The choice of space of rays rather     
than the Hilbert space is in accordance with the quantum mechanical            
arbitrariness of phase of the state vectors. The finite distance  
${\cal D}(\overline{\ket{\z_1}},\ \overline{\ket{\z_2}})$ between any two rays
$\overline{\ket{\z_1}}$ and $\overline{\ket{\z_2}}$, which are 
associated with the
normalized                                                                      
coherent states $e^{i\varphi_1}\ket{\z_1}$ and $e^{i\varphi_2}\ket{\z_2}$, is   
defined by                                                                        
\be                                                                            
{\cal D}^2 (\overline{\ket{\z_1}},\ \overline{\ket{\z_2}})=
\inf_{\varphi_1,\varphi_2} ||           
e^{i\varphi_1}|\z_1\rangle-e^{i\varphi_2}|\z_2\rangle ||^2=
2-2\abs{\inner{\z_1}{\z_2}}\;.
\label{dist}                                                        
\ee
                                                                            
\noindent This is a proper distance function as it is positive definite and 
non-degenerate, 
it satisfies the triangular property and it is also gauge invariant. In addition    
its infinitesimal form gives the distance of two nearby coherent state vectors  
and provides the metric tensor on the CS-manifold~\cite{de}, \ie
\bea
(ds)^2&=&{\fscr{1}{2}}{\cal D}^2 (\overline{\ket{\z+d\z}}\ ,\ 
\overline{\ket{\z}})=
1- \abs{\inner{\z + d\z}{\z}} \nn \\             
&=&\left[\frac{\partial_{\z}\partial_{\bar{\z}}\; (\z|| ||\z)}{(\z|| ||\z)}-       
\frac{\partial_{\bar{\z}}\;(\z|| ||\z)}{(\z|| ||\z)}\cdot
\frac{\partial_{\z}\;(\z|| ||\z)}
{(\z|| ||\z)}\right] d\z d\bar{\z} \nn \\  
&=&  \frac{\partial}{\partial \absqr{\z}}\left[
\frac{\absqr{\z}}{(\z|| ||\z )}\, \frac{\partial (\z || ||\z )}
{\partial \absqr{\z}}
\right]
d\z d{\bar \z} \equiv g_{\z\bar{\z}} d\z d\bar{\z},
\label{line}
\eea

\noindent or alternatively
\be                                                                             
(ds)^2=\left[\frac{(\z|| G_- G_+ ||\z)}{(\z|| ||\z)}-\frac{(\z|| G_-||\z)}
{(\z|| ||\z)}
\cdot\frac{(\z|| G_+ ||\z)}{(\z|| ||\z)}\right] d\z d\bar{\z}\ .    
%\label{19}                             
\ee       
                                                                      
\noindent The curvature scalar~\cite{h},                                      
\be
R=-g^{-1}_{\z\bar{\z}}\partial_\z\partial_{\bar{\z}}(\ln
g_{\z\bar{\z}}),
%\label{20}
\ee       
                                                                      
\noindent following from the metric $g$ involves higher correlations of the 
$G_\pm$
generators and is in general not constant. 
It is worth noticing that the basic geometric objects of both the
symplectic and Riemannian structure endowed in $\cal M$ are given in 
terms of the so called symbols of the operators or their products~\cite{ber},
\ie  the
coherent state mean value 
$\bra{\z}{\rm G}\ket{\z} \equiv \braket{\rm G}$ of the 
corresponding operator $\rm G$. Also the noncommutativity of the involved 
operators and their non-zero uncertainties in the coherent state basis is
essential for the non-trivial geometric characteristics of the $\cal M$
manifold. In effect, the CS-manifold $\cal M$ captures some genuine 
quantum mechanical features despite its classical character; this property
of $\cal M$ is further manifested in the quantum mechanical evolution to
be studied shortly.

\nd As the models to be studied in the next sections are of boson
and fermionic/spin type with dynamical symmetries related to the WH and 
the $su(2)$ algebras~\cite{gil}, we now exemplify the above construction for the
CS of these algebras. 
\nd For the boson algebra,
\be
[a,\ad ]=\bf{1} \ \ \ \it{[N,\ad ]=\ad \ \ \ [N,a ]=-a \;,}
\ee

\nd with $N=\ad a$ the number operator, the vacuum state is the zero-photon
state vector $\ket{\oslash}=\ket{0}$, and ${\rm G_+}=\ad$ the creation operator.
This defines the boson CS
\be
\ket{\a}=e^{\a\ad-\ab a}\ket{0}=
{\cal N} e^{\a\ad}\ket{0}=e^{-{\fscr{1}{2}}\absqr{\a}}\sum_{n=0}^{\infty}
\frac{\a^n}{\sqrt{n!}}\ket{n}\;.
\ee

\nd It is an (over)complete set of states with respect to the measure
$\mes{\a}=\frac{1}{\pi}e^{-\absqr{\a}} d^{2}\a$ for the non-normalized CS, 
and $\a\in{\cal M}=
WH/U(1)\approx C$ 
is the CS manifold. Since $a\ket{\a}=\a\ket{\a}$,
which implies the symbols $\braket{a}=\a$, $\braket{\ad}=\ab$, $\braket{N}=
\absqr{\a}$, $\cal M$ is the flat canonical phase plane with the
standard line element $ds^2 =d\a d\ab$. Also the symplectic 2-form 
$\om=id\a \wedge d\ab$ is associated to the canonical Poisson bracket
$\{f,g\}=i(\pl{\a}f\pl{\ab}g - \pl{\ab}f\pl{\a}g) $.

Next we come to the case of the $su(2)$ algebra with commutation 
relations,
\be
[J_0 , J_\pm ]=\pm J_\pm  \ \ \ [J_+ , J_{-} ]=2 J_0 \;.
\ee

\nd The vacuum state $\ket{\oslash}\equiv\ket{j\ -j}$ is the extremal vector in 
the representation module of the algebra, with $\{ \ket{j\ m}\}_{m=-j}^{j}$, 
$j=1/2, 1, 3/2, \ldots$, and
$-j\leq m\leq j$. 
The CS vector is obtained by displacing the vacuum with a $SU(2)/U(1)$ coset
element:
\be
\ket{z}=e^{\nu J_+ - \nub J_-}\ket{j\ -j}={\cal N}e^{zJ_+}\ket{j\ -j}=
{\cal N}\sum_{m=-j}^{j}\col{2j}{j+m}^{\fscr{1}{2}} z^{j+m} \ket{j\ m}\;,
\ee

\nd with $z=(\nu /|\nu|){\rm tan}|\nu|$ and normalization coefficient 
${\cal N}=(1+z\zb)^{-j}$. The CS manifold
${\cal M}=SU(2)/U(1) \approx {\rm S^{2}} \approx  {\rm CP^1}$ 
is isomorphic
to the
2-sphere $\rm S^2$, with $0\leq\th\leq\pi \;, 0\leq\varphi\leq2\pi$ polar and 
azimuthal
angles respectively. It is also parametrized as the complex 
projection plane $\rm CP^1$ with coordinate 
${\cal M}\ni z={\rm tan}({\fscr{1}{2}}\th)e^{-i\varphi}$.

\nd The symbols of the generators are 
\bea
\label{symb}
\braket{J_+}&=&2j\frac{\zb}{1+z \zb} \; \; \; , \; \; \;
\braket{J_-}=2j\frac{z}{1+z \zb} \nn \\
\braket{J_0}&=&-j\frac{1-z\zb}{1+z \zb} \; \; \; , \; \; \; 
\braket{J_- J_+}=\frac{4j^2 z\zb +2j}{(1+z \zb)^2} \;.
\eea

\nd Based on the general scheme the above generator
symbols give rise to the line element $ds^2 =(1+z\zb)^{-2} dz d\zb$ 
for the complex tangent plane $\cal M$. The (over)completeness of 
the non-normalized CS  invokes the Haar invariant measure on $\cal M$,
$\mes{z}=\frac{2j+1}{\pi}(1+z\zb)^{-2}d^2 z$. The CS vector space is now the
constant curvature ($R=1$) sphere equipped with the canonical 
1-form $\theta = ij \frac{\zb}{1+z\zb} dz + \rm{cc}$, which 
furnishes a canonical
symplectic 2-form $\om=i j\frac{dz \wedge d\zb}{(1+z \zb)^2}$, and the 
associated 
Poisson bracket becomes 
$\{f,g\}=\frac{i(1+z\zb)^2}{2j}(\pl{z}f \pl{\zb}g - \pl{\zb}f \pl{z}g)$. 

\section{ Generalized discrete self-trapping equation }
The classical Hamiltonian of the generalized discrete self-trapping (GDST)  
model reads~\cite{els,cs} 
\be
H_{\rm \scriptstyle CL} = \sum_{j=1}^f (\omega_0 |A_j|^2 - \frac{\gamma}{m} 
|A_j|^{2m})
- \sum_{j \neq k}^{f} \l_{jk} \overline{A}_j A_k \;,
\ee
\noindent where $j=1,2,\ldots,f$ counts the number of complex mode amplitudes 
$A_j(t)$ and
its complex conjugate $\overline{A}_j(t)$. The equation of motion
\be
i\dot{A}_j=\omega_0 A_j - \sum_{(j \neq k) k=1}^{f} \l_{jk} A_k - \gamma |A_j| 
^{2(m-1)}A_j
\label{GDST}
\ee
\noindent is derived from the canonical Poisson brackets
\be
\{A_j,\overline{A}_k\}=i\delta_{jk}  , \;\;\;  
\{A_j,A_k\}=\{\overline{A}_j,\overline{A}_k\}=0
\ee
\noindent using the equation of motion $\dot{A}_j=\{H,A_j\}.$

\noindent The quantization of this canonical Lie-Poisson algebra proceeds 
with the correspondance
rule~\cite{dir,weyl} $A \longrightarrow b, A^\ast \longrightarrow b^\dagger$, 
where 
$b^\dagger$ and $b$ are 
the canonical creation and annihilation operators respectively.
To account for problems of ordering, well known
in the naive quantization rule, we shall here confine ourselves  to 
two ordering rules, namely the normal ordering (NO) and the 
symmetric ordering (SO). SO is a classically motivated symmetrization of 
noncommuting operators and should be expected to give the right classical limit
when $\hbar \rightarrow 0$. However here we study the quasiclassical limit
of the quantum equations of motion by means of the CS 
symbol of operators and in this case the NO is the appropriate ordering.
(For further discussion on the difference between semi- and quasiclassical
ordering see refs. 89 and 142 in~\cite{kla}, ff p.64-65).
 
\nd For the SO quantization rule the correspondence~\cite{hs,chjs} 
\begin{equation}
|A|^{2m} \longrightarrow \frac{m!}{2^m} L_m (-2x) \;,
\end{equation}
\noindent is valid for quantization of arbitrary positive powers of the
modulus of the complex amplitude;
here $x^k=\::(b^{\dagger}b)^k:\: \equiv b^{\dagger k}b^k$ 
and $L_m(\cdot)$ is the Laguerre polynomial of zero order.
Then, using the symmetric ordering 
quantization rule and after a constant shifting by 
$\frac{1}{2}\omega_0-\g \frac{(m-1)!}{2^m}$,
the Hamiltonian of the GDST model takes the form
\begin{equation}
H_{\rm \scriptstyle SO}=H_{\rm \scriptstyle NO} - 
\g \sum_{j=1}^{f}\sum_{n=1}^{m-1} \m_{n}^{(m)} {b^\dagger_j}^n  {b_j}^n\;,
\label{hamso}
\end{equation}

\nd where $\m_{n}^{(m)} =\frac{(m-1)!}{2^m} 
\left( \begin{array}{c}
m\\
m-n
\end{array} \right) \frac{2^n}{n!}$, and
\begin{equation}
H_{\rm \scriptstyle NO}=\sum_{j=1}^{f} (\omega_0 b^\dagger_j b_j - 
\frac{\gamma}{m} b^{\dagger m}_j
b^m_j) - \sum_{j \neq k}^{f} \l_{jk} b^\dagger_j b_k \;,
\label{hamno}
\end{equation}
is the Hamiltonian obtained by the normal ordering quantization rule.
We observe that the symmetric ordering rule adds $m-1$ additional terms in
the Hamiltonian with respect to its NO quantization.
However since for 
$m=2$ 
it is reduced to the number operator which can be absorbed in the 
$H_{\rm \scriptstyle NO}$ 
by simply
redefining the coefficient of the corresponding number operator term, 
we see that in effect the symmetric ordering modifies the Hamiltonian
only for $m>2$, \ie  only for higher order nonlinearity.

Next we obtain the Heisenberg equations of motion for the boson operators
by using the Hamiltonian of eq.~(\ref{hamso})
\be
i \dot{b}_j=\omega_0 b_j-\g b^{\dagger m-1}_j
b^m_j - \sum_{k=1}^{f} \l_{jk} b_k 
- \g (\sum_{n=1}^{m-1} \m_{n}^{(m)} n {b^\dagger_j}^{n-1}  {b_j}^n)\;.
\label{hequ}
\ee

\nd Making use of the factorized state vector 
$|\beta> \equiv \otimes_{j=1}^{f} |\beta_j>$, where
$b_j |\beta_j> = \beta_j |\beta_j>, j=1,...,f$, we evaluate the CS mean value
of the last equation to obtain
\be
i\dot{\beta}_j=
\omega_0 \beta_j - \g 
|\beta_j|^{2m-2} \beta_j - \sum_{(j\neq k) k=1}^
{f} \l_{jk} \beta_k - \g \sum_{n=1}^{m-1} 
\m_{n}^{(m)} n |\beta_j|^{2n-2} \beta_j\;.
\label{eqcite}
\ee

\nd By means of the canonical Poisson brackets this set of equations 
can be derived from the Hamiltonian
\be
{\cal H}=\sand{\b}{H}{\b}=
\sum_{k=1}^{f}(\omega_0|\beta_k|^2 - \frac{\gamma}{m}|\beta_k|^{2m}) -
\sum_{ k \neq l}^{f} \l_{kl} \beta_k \overline\beta_l - 
\g \sum_{k=1}^{f}\sum_{n=1}^{m-1} 
 \m_{n}^{(m)}  |\beta_k|^{2n}.
\label{ham}
\ee

\noindent The above Hamiltonian and the norm $N=\sum_{ j=1}^
{f} |\beta_j|^2 $ are the constants of motion for the dynamics issued
by eq.~(\ref{eqcite}). We observe that the first two sums in 
eq.~(\ref{ham}) combine to give 
the coherent state symbol of the $H_{\rm \scriptstyle NO}$ of 
eq.~(\ref{hamno}), while the 
remaining part 
corresponds to the symbol of the terms appearing 
in eq.~(\ref{hamso}) due
to the SO. This implies that NO quantization induces in the CS manifold 
$\cal M$ an
essentially classical time evolution while the quantization by SO, 
due to the 
additional terms, induces a dynamical evolution on $\cal M$ which departs from
classical dynamics. This of course makes a difference in the dynamics of the 
observables
of the system as well as in the wavefunction distributions.
At this point we should remark that the assumption of factorization of the 
state vector into CSVs implies that any normal ordered function of boson
operators from each mode has CS mean value expressed by the same
function, {\it i.e.}
\be
\sand{\a}{:f(\ad_{j}, a_{j}):}{\a}=f(\ab_{j}\,,\a_j)\,.
\ee

\nd Also for each oscillator mode the Hermitian combinations 
$x=\frac{1}{\sqrt 2}
(\ad+a)$ and $p=\frac{i}{\sqrt 2}(\ad-a)$, {\it i.e.} the position and momentum
operators, we obtain the minimum uncertainty property of the boson CSV,
\be
\D x\D p=\frac{1}{2}\,,
\label{unpr}
\ee

\nd for the uncertainties $(\D x)^2=\braket{(x-\braket{x})^2}$ and
$(\D p)^2=\braket{(p-\braket{p})^2}$. We shall recall this property shortly
for the Q-function. 

\nd To illustrate these differences in dynamics due to ordering 
we shall employ the quantum 
mechanical Q-distribution function~\cite{qfun} and the Poissonian 
distribution of the boson excitations of a CSV in the number state 
basis~\cite{gla}. For the Q-distribution of a $\rho$ density matrix
we define 
\be
Q_j=\frac{1}{\pi} \bra{\beta_r}\rho_j\ket{\beta_r}
=\frac{1}{\pi}\absqr{\inner{\beta_r}{\beta_j}}\,,
\label{qfdef}
\ee

\nd with $j=1,\ldots,f$. Here we have assumed a pure state for each
mode \ie
$\rho_j=
\proj{\beta_j}{\beta_j}$, and the reference CS $\ket{\beta_r}$ is labelled
by the complex coordinate $\beta_r=x_r +iy_r$. The Q-function then takes the 
form
of a displaced Gaussian function \viz
\be
Q_j(x_r,y_r)=\frac{1}{\pi}e^{-(x_{r} -{\rm Re}\beta_j)^2 -(y_{r} -{\rm 
Im}\beta_j)^2} \,.
\ee

\nd The center of the Gaussian $({\rm Re}\b_{j} , {\rm Im}\b_{j})$ is the 
expectation value of the position and momentum operators in the CS basis, 
{\it i.e.} $({\rm Re}\b_{j} , {\rm Im}\b_{j})=(\sand{\b}{x_j}{\b},
\sand{\b}{p_j}{\b})$, for $j=1,\ldots,f$. However there is a
nonzero quantum mechanical uncertainty around this point since 
$\D x_{j}=\D p_{j}=\frac{1}{\sqrt 2}$, 
albeit it is the minimum acceptable one (\cf.~eq.(\ref{unpr})).

In Figs. 1(a,b) the Gaussian bells of the Q-functions for the 
quintic $(m=3)$ GDST
trimer $(f=3)$ system are plotted at a certain instant of time for SO and
NO respectively. Here and in the following we use periodic boundary
conditions and choose $\l_{jk}=1$ 
for nearest neighbouring sites, and zero
otherwise. The centers of the Gaussian functions of Fig. 1 are plotted
in Fig. 2 for the SO (Fig. 2a,b) and the NO (Fig. 2 c,d) cases,
respectively. We note
that in the SO case the trajectory of the bell corresponding to the
initially excited site will always be confined in a region far from
the origin, while the bells corresponding to the other two sites will
stay close to the origin. In our framework, this corresponds
to the well known self-trapping of the classical system. It is obvious
that the appearance of the extra terms in the SO-equation enhances the
self-trapping compared to the NO-case.

\nd Next we consider the Poissonian distribution of each boson mode in the
coherent state,
\be
P_{n}^{j}=\absqr{\inner{\beta_j}{n}}=
e^{-\abs{\beta_j}^2}\frac{\abs{\beta_j}^{2n}}
{n!}\,,
\label{poiss}
\ee

\nd where $n=0,1,\ldots$ enumerates the number states and 
$\absqr{\beta_j}=\sand{\beta_j}{b^{\dagger}_j b_j}{\beta_j}$ is the 
expectation value of the number operator  which gives the 
average number of quanta in the CS basis.
An illustration of how the
self-trapping reflects itself in the Poissonian distribution and how 
it gets enhanced in SO quantization, is shown
in Fig. 3a and 3b for the SO and NO cases respectively. The
system is here a rather large $(f=21)$ quintic $(m=3)$ GDST-system, and
initially one single site $(j=21)$ is excited with the total excitation
number $N=10$. As can be seen, for the particular parameter values
chosen here the main part of the excitation stays at the initially
excited site in the SO-case, while it spreads more or less equally
among the modes in the NO case.

To end this section, we note that in the simple case of the GDST dimer
$(f=2)$ with cubic $(m=2)$ or quintic $(m=3)$ nonlinearity, it is
possible to derive an analytic expression for the critical
value,$\g_{\rm \scriptstyle cr}$, of the 
nonlinear coupling coefficient for the appearance of self-trapping in
eq.~(\ref{eqcite}). This is done in the standard way by considering
the variable $r=|\beta_1|^2-|\beta_2|^2$. For the ordinary GDST
equation~(\ref{GDST}), it was found in ref.~\cite{kenkre} for the
cubic case and in ref.~\cite{cs} for the quintic case, that $\ddot{r}$
could be expressed as a cubic polynomial in $r$, and solutions in
terms of Jacobi elliptic functions were obtained. In the case of
eq.~(\ref{eqcite}), containing both cubic and quintic nonlinearity, we
find the same equation for $\ddot{r}$ as in ref.~\cite{kenkre}, provided
that we replace $\gamma$ with $\gamma^{(2)}+N\gamma^{(3)}$,
where $\gamma^{(2)}$ and $\gamma^{(3)}$ are the
coefficients of the cubic and quintic nonlinear terms,
respectively. Thus, defining $\g_{\rm \scriptstyle cr}$ in the usual
way as the smallest value of $\g$ for which $\absqr{\b_{j_{0}}}$
is always
larger than $N/2$, given that the excitation initially was localized on
site $j_0$, we find  the self-trapping condition 
$\frac{\gamma^{(2)}+N\gamma^{(3)}}{\l_{jk}} > \frac{4}{N}$ from
  the explicit solution for $r(t)$ just as in ref.~\cite{kenkre}. Using
  the explicit expressions for $\gamma^{(2)}$ and $\gamma^{(3)}$ 
in ~(\ref{eqcite}) and
  $\l_{jk}=1$, the following expressions for 
$\g_{\rm \scriptstyle cr}$ are
obtained in the NO and SO cases, respectively: 
\be
\g_{\rm \scriptstyle cr}^{\rm \scriptstyle NO}=\frac{4}{N^2} \ \ \ \;\;, \ \ \
\g_{\rm \scriptstyle cr}^{\rm \scriptstyle SO}=\frac{4}{N(N+3)} \;.
\label{gcr}
\ee
These expressions are plotted in Fig. 4 as a function of the 
total number of lattice quanta $N$. In the limit of large number 
of excitations (phonons)
the quantum lattice model, according to Bohr's correspondance principle, will
behave classically. In particular the critical value of $\g$ for the onset
of self-trapping, which as we have also seen from previous plots is different
for SO and NO, is expected to become equal in the classical limit of the
quantum lattice models. 
In this limit the ordering of the
operators, a genuine quantum characteristic, should not be of
importance.
Indeed the analytic expressions~(\ref{gcr}) 
for the critical $\g$'s of the  GDST
dimer show, that when the number 
of quanta becomes large, the phenomenon of DST occurs
asymptotically for the
same values \ie  $ \g_{\rm \scriptstyle cr}^{\rm \scriptstyle NO}\approx
\g_{\rm \scriptstyle cr}^{\rm \scriptstyle SO}$, independently of the ordering
rule.

\section{Validity of the factorization ansatz and quantum correlations}
Let a composite quantum system be described by a general pure state
$\ket{\psi}$ living in ${\cal H} =\otimes _{i=1}^{f}{\cal H}_{i}$,
the tensor product of Hilbert spaces of its $f$ subsystems.
In general the form of such a generic vector $\ket{\psi}$ can be
of three different types, each one signifying a specific kind of
correlation among the subsystems. For vectors of the product
form, $\ket{\psi}=\ket{\phi}^1 \otimes \ket{\phi}^2 \otimes 
\ldots\otimes \ket{\phi}^f $,
the subsystems are decorrelated. If $\ket{\psi}$ is a convex combination
of product states \ie 
$\ket{\psi}=\sum _{i}^{n} \lambda _{i}\ket{\phi}_{i}^{1} 
\otimes \ket{\phi}_{i}^{2} \otimes
\ldots \otimes \ket{\phi}_{i}^{f} $, with $\sum _{i}^{n} \lambda _{i} =1$,
the subsystems are said to be classically correlated~\cite{peres}.
For inseparable or entangled systems, the third case, the state
vector $\ket{\psi}$ is a general superposition of states from its subsystems
with no additional properties, that are further characterized by \eg violation 
of 
Bell inequalities~\cite{bell}; this is the case when genuine quantum 
correlations are developed among the subsystems.

Although the existence and the oddities of quantum nonlocal correlations and of
entanglement have been pointed out since the early days of Quantum Mechanics
\cite{epr}, and have remained an active subject of research since then
\cite{bell}, it is only recently (due mainly to some striking
applications of entanglement such as quantum computation, cryptography,
teleportation etc; see \cite{D.entang,ef} and refs. therein ), that measures
that quantify the amount of correlations of a given state of a composite 
quantum system have been studied. In the above framework
of quantum correlations it is clear that our factorization ansatz
is equivalent to the assumption that no quantum correlations will be
developed in the course of time evolution. Still, as we have seen, due to
the nonlinear form of the quantum Hamiltonian the equations of motions 
derived from that approximate ansatz differ from the entirely classical ones.
This difference captures some of the quantum features of the system, so
we would like to have a figure of merit of our approximation. To this end
we introduce a single index, the correlation index, that measures the distance 
of the 
true state of the model from the factorized one that is assumed, and so gauges 
the validity 
of our approximation. 

Recall that the CS-vectors form a basis for the Hilbert space of a single site, 
so for the $f$ sites of the boson chain model a general state vector can 
be expanded as 
\be
\ket{\psi}=\int P(\{\a _i \} , \{ \ab_i \} ) \otimes _{i=1}^{f}\ket{\a _i}
d\mu (\{\a _i \})
\label{expan}
\ee

\nd with $d\mu (\{\a _i \})=\prod _{i=1}^{f} d^2 \a _i$. By means of the
relations~(\cite{louis}),
\bea
f(\bar{\b}) &=& \int f(\ab ) \exp [ {-\half (\absqr{\a} + \absqr{\b} ) 
+\bar{\b} \a } ] d^2 \a \;,\\ \nn
f(\bar{\b} ) &=& \int f(\ab ) \inner{\b }{\a }d^2 \a \, ,
\eea

\nd which imply that the overlap $\inner{\b }{\a }$ of CS behaves as a delta 
function 
for the integrable functions of the Bargmann-Hilbert space, we can compute
that  $\ket{\psi}$ is normalized provided that $\int \absqr 
{P(\{\a _i \} , \{ \ab_i \} )} d\mu (\{\a _i \})=1. $
This shows that if $P(\{\a _i \} , \{ \ab _i \})= \d (\a _1 - \a_1 ^r ) \dots
\d (\a _f - \a_f ^r )$, then the general vector $\ket{\psi}$ is reduced to the 
factorized state $\ket{\psi _{ref}}= \otimes _{i=1}^{f} \ket{\a _i ^r}.$

\nd We now introduce the correlation index $\e$ (\cf eq. (\ref{dist}) ),
\bea
\e &=& {\cal D}^2 (\overline{\ket{\psi}},\ \overline{\ket{\psi _{ref}}})=
2-2\abs{\inner{\psi }{\psi _{ref}} }
\\ \nn 
&=& 2-2 \abs{\int d\mu (\{\a _i \}) P( \{\a _i \} , \{ \ab _{i} \} )
\prod _{i=1}^{f}\exp [ {-\half (\absqr{\a _i ^r} + \absqr{\a _i} ) +
\ab _{i}^{r} \a _{i} }] }.
\eea

\nd Initially $\e (t=0)=0$, since we set $\ket{\psi(t=0)}=\ket{\psi _{ref}}$.
In any subsequent time interval $a\leq t \leq b$, for which $\e (t=0)\approx 0$,
our ansatz is justified. To determine $\e (t)$, we need to know the kernel
$ P(\{\a _i \} , \{ \ab _i \} )$. Substituting the expansion~(\ref{expan})
in the Schr\" odinger equation we find that the kernel $P$ satisfies
the Fokker-Planck equation,
\be
i\hbar  \partial _t P(\{\a _i \} , \{ \ab _i \} )= 
{\cal H}(\a _i , -\partial _{\a_ i}  + \half \ab _{i} ) 
P(\{\a _i \} , \{ \ab _i\} ) \, ,
\ee

\nd where ${\cal H}$ is the analytic image of the Hamiltonian operator under
the substitutions $a_i \rightarrow \a _i \, , \, \ad _i \rightarrow
-\bar{\partial} _{\a _i} $. 
There are a number of approximate and analytic solution 
techniques available for such equations~\cite{risk},
most accessible \eg is the technique of continued fractions.
However the problem of quantum correlations in nonlinear lattice models is 
interesting even in the dimer case, and as far as we know it has not been 
addressed.
In such a case the conservation of the total number of excitation quanta
in the lattice sites will make the number of terms in a continued
fraction method of solving the ensuing Fokker-Planck equation finite, since the 
dimensionality of the available Hilbert space
and of the $\rho$-density matrix is determined by the number of quanta 
available.
This makes possible an explicit calculation of the correlation index
and allows to set a number of interesting questions concerning  
the relation of correlation/decorrelation versus selftrapping/non-selftrapping
in the case of the quantum dimer (see \eg ~\cite{ellman} for a full quantum 
treatment
of the quantum dimer); however more work should be 
anticipated along these lines~\cite{ejc}. Finally, we note that for
non-boson lattices, as the ones of the last appendix, the introduction of
the correlation index can be done again based on the distance function
defined for the respective CSV, \cf eq.(\ref{dist}). 

\section{Conclusions}
This work was inspired by the fact that for a linear quantum
mechanical oscillator there is a coherent state vector defined in its 
Hilbert space, predicting that the motion will be centered around the
classical path with the minimum uncertainty. Here, this simple idea
has been extended to some quantum nonlinear chain models.
For quantum chain models with an oscillator attached at each site, the
obvious assumption which has been followed here
is that each site is in a CS ; the same 
idea however can be applied for multispin-like models with some advantages
as we have shown. The essential geometric nature of our assumption
has also been stated in terms of the symplectic and Riemannian geometry 
of the CSV spaces. Other types of special states, such as \eg even/odd
CSV (symmetric/antisymmetric combination of boson CS) which are
associated with different geometries of their CS spaces, could have been
considered; the final choice depends on the dynamical symmetry of the given 
Hamiltonian and on the specific form of the nonlinear 
interaction terms. 

Also, the treatment of the DST in the quantum regime
as has been presented here shows that this is a phenomenon which continues
to exist after quantization of the classical DST equations. However, 
as has been demonstrated by means of the CS method, its
exact parameter dependence is crucially affected by the quantization
scheme. Contrary to the NO case where the situation is essentially classical,
in the SO rule DST gets even more pronounced
from its classical form. We note that a similar statement can be made
also concerning the Hartree approximation treated
in~\cite{wehms,hs,seg}, since this ansatz results into similar
approximate dynamics as obtained here. This can be seen by comparing
our eq.~(\ref{eqcite}) with eqs. (4.3) and (4.4) of
ref.~\cite{hs}. However, the Hartree ansatz is basically different
from the factorization ansatz used here. Namely, geometrically it can
be said that the two ans{\"a}tze confine the dynamics in different
subspaces of the total Hilbert space of the GDST model, moreover
the Hartree wavefunction is not fully factorized.

Finally, we should note that although the factorization hypothesis for
the state vector has been effective in treating the dynamical
evolution in the model-cases studied here, its validity should be
questioned. This is a crucial point, since factoring the state vector
of a compound quantum system implies a loss of the quantum
correlations which may be developed during the course of the time
evolution. The failure of a factorized state to account for such
entanglement of quantum subsystems can however be
quantified
by means of a correlation index, as we have shown. In future work we
plan to present a study  of the entanglement phenomenon versus DST in a
quantum dimer model.

\section{Acknowledgements}
We thank A. C. Scott for valuable comments. One of us (D. E.) acknowledges
support from the EU HCM programme under grant
No. CHRX-CT93-0331, he is also grateful to the Institute of Mathematical 
Modelling (IMM) for hospitality and to G. P. Tsironis for discussions.
M. J. acknowledges financial support from the
Swedish Foundation for International Cooperation in Research and
Higher Education. Constructive comments by the referee are also acknowledged. 

\section*{Appendix 1: Modified discrete nonlinear Schr\"odinger equation}
The ability of the CS-method to extract the quantum mechanical 
characteristics of the time
evolution for a given lattice Hamiltonian model is very much depending on
the special form of the Hamiltonian itself. Occasionally the derived 
 equations of motion will be reducible to the classical ones
indifferently of the quantization ordering rule (at least for the common
rules); let us take up the 
following case. 
The so called modified discrete nonlinear Schr\"odinger equation 
(MDNLS)~\cite{kto} has been
recently introduced as a model describing in the adiabatic limit 
polaron dynamics
when the excitation is coupled to an acoustic chain of oscillators. Compared to 
ordinary DNLS dynamics of an excitation coupled to optical oscillators, 
the MDNLS contains some new formal features such as off-diagonal 
nonlinearity, and for a ring with $f$-sites 
reads
\begin{equation}
i\dot{A}_j=V(A_{j+1}+A_{j-1}) - X (|A_{j+1}|^2+|A_{j-1}|^2+2|A_j|^2)A_j \;.
\end{equation}

\noindent Using canonical Poisson brackets, these equations can be derived 
from the 
Hamiltonian
\begin{equation}
H_{\rm \scriptstyle CL}=V\sum_{j=1}^{f}(A_j \overline{A}_{j+1} 
+ \overline{A}_j A_{j+1}) -X\sum_{j=1}^
{f} (|A_{j+1}|^2 |A_{j+2}|^2 + |A_j|^4).
\end{equation}
\noindent The quantization of the classical Hamiltonian for the MDNLS 
equation leads (after a constant shifting of the Hamiltonian) to the 
following expression
\begin{equation}
H_{\rm \scriptstyle SO}=H_{\rm \scriptstyle NO}-{\fscr{1}{2}} X\sum_{j=1}^{f} 
(b_{j+2}^{\dagger} b_{j+2} + b_{j+1}^{\dagger}
b_{j+1} + b_j^\dagger b_j)\;,
\label{mham}
\end{equation}
where 
\begin{equation}
H_{\rm \scriptstyle NO} = 
V \sum_{j=1}^{f} (b_j b_{j+1}^\dagger + b_j^\dagger b_{j+1}) - X 
\sum_{j=1}^{f} (b_{j+1}^\dagger b_{j+1} b_{j+2}^\dagger b_{j+2} 
+ b_j^{\dagger^2}  
b_j^2)\;.
\end{equation}

\nd The first term in eq. (\ref{mham}) refers to the NO and the second one 
is obtained as before from the SO rule. Since the total Hamiltonian is 
written in terms of WH algebra elements defined at each site and by their 
products,
we can follow exactly the same procedure as in the quantum DNLS case. 
Namely we assume the factorization of the state vector 
in terms of canonical CSVs. By the method of the previous section, 
and for the Hamiltonian $H_{\rm \scriptstyle SO}$, this leads to 
derivation of a set of 
quasiclassical equations of motion satisfied by the CSV parameters $A_j$,
\begin{equation}
i\dot{A}_j=V(A_{j+1}+A_{j-1}) - X (|A_{j+1}|^2+|A_{j-1}|^2+
2|A_j|^2)A_j +{\frac{3}{2}}X A_{j}\,.
\end{equation}

\nd The above expression shows that the quasiclassical equation differs by a 
constant shift of the coefficients of the onsite variables from the 
corresponding classical equation. With the transformation 
$A_{j}\rightarrow e^{-{\fscr{3}{2}}iXt}A_j$ 
they become the classical equations.
The formal similarity of classical and quasiclassical equations
allows to conclude the following:
the particular form of the MDNLS Hamiltonian after quantization by NO and SO
generates, (on the geometrical space singled out by the factorization of the 
model's quantum state in terms of boson CSV), an (almost) classical
dynamics.
This however as the previous sections show, should not be expected to be
true in general.

\section*{Appendix 2: Fermionic polaron model (FP)}
To demonstrate the ability of the proposed method to study 
quantum dynamics for models
others than quantized boson chains, we shall take up in this appendix the 
fermionic polaron (FP) model. It has been recently proposed to describe the
interaction of electrons with optical phonons in one spatial 
dimension~\cite{fed}. This same model has been studied from the point of view
of QISM~\cite{pz,esse} and the NSM~\cite{esse,seg}.
Under the 
assumptions made in~\cite{fed} the Hamiltonian of the 
interacting
electron-phonon system with periodic boundary conditions reads
\begin{equation}
H=\epsilon\sum_{j=1}^{N}a_j^\dagger a_j - g\sum_{j=1}^{N} (a_j^\dagger
a_{j+1} + a_{j+1}^\dagger a_j) + V
\sum_{j=1}^{N} n_j n_{j+1} \; ,
\label{hamfp}
\end{equation}

\noindent where the fermion annihilation operator of a polaron at lattice site
$j$, $a_j$ and its corresponding Hermitian conjugate $a^\dagger_j$
(the creation operator)
generate the Grassmann algebra
\begin{equation}
[a_j ,a_k ]_+ =[a_j^\dagger, a_k^\dagger ]_+ =0 \; \; \;,\; \; \;
[a_j ,a_{k}^{\dagger} ]_+ =\delta_{jk} \;.
\label{gras}
\end{equation}

\noindent The occupation number operator is defined by $n_j=a_j^\dagger a_j$,
while  $g$ is a parameter proportional to the overlapping integral 
and $V$ 
stands for the electron-phonon coupling constant (see \cite{fed} for details).
Using the equivalence of the FP model to the XXZ model provided in~\cite{pz} 
 by means
of a Jordan-Wigner transformation, we shall proceed to write 
the spin analogue of
the fermionic model. The Grassmann algebra in eq.~(\ref{gras}) is
realized in terms of a
tensor product of 
generators of the (spin $1/2$ Pauli) algebra 
by means of the following Jordan-Wigner transformation
\bea
a_j&=&\sm_j {\rm exp}\left(i\pi \sum_{k=1}^{j-1}
(\sigma_k^z + {\fscr{1}{2}} {\bf{1}}_j)\right) \nn \\
a_j^\dagger&=&\sp_j {\rm exp}\left(-i\pi \sum_{k=1}^{j-1}
(\sigma_k^z + {\fscr{1}{2}} {\bf{1}}_j)\right) \nn \\
n_j&=&\sigma_j^+ \sigma_j^-=({\fscr{1}{2}} {\bf{1}}_j +\sigma_j^z)\;,
\label{jw}
\eea

\noindent and its inverse
\be
\sigma_j^-=a_j {\rm exp}(- i\pi \sum_{k=1}^{j-1} n_k)    \ \ \ \;,\; \ \ \
\sigma_j^+=a_j^\dagger {\rm exp}(i\pi \sum_{k=1}^{j-1} n_k)  \nn 
\ee

\be
\sigma_j^z=n_j - {\fscr{1}{2}} {\bf{1}}_j \;,
\ee

\nd where $\sigma^{\pm}_j = \sigma^{x}_j \pm i \sigma_j^y$, and $\sigma_j^x,
\sigma_j^y, \sigma_j^z$ are the Pauli matrices and ${\bf{1}}_j$ is the
unit matrix at site $j$.

\noindent Using eq.~(\ref{jw}) the transformed Hamiltonian of eq.~(\ref{hamfp})
reads,
\be
H=\frac{N}{2}(\epsilon + \frac{V}{2}) + (\epsilon
+ V) \sum_{j=1}^{N} \sigma_j^z + H_{\rm \scriptstyle XXZ} \;,
\ee

\noindent where
\be
H_{\rm \scriptstyle XXZ}=V \sum_{j=1}^{N} \sigma_j^z \sigma_{j + 1}^z -
2g \sum_{j=1}^{N} (\sigma_j^x \sigma_{j + 1}^x +
\sigma_j^y \sigma_{j + 1}^{y}) \;,
\ee

\noindent or
\be
H_{\rm \scriptstyle XXZ}=
V \sum_{j=1}^{N} \sigma_j^z \sigma_{j + 1}^z
- g \sum_{j=1}^{N} (\sigma_j^+ \sigma_{j + 1}^{-} + \sigma_j^-
\sigma_{j + 1}^+ )\;.
\ee

\noindent We now proceed with the $H_{\rm \scriptstyle XXZ}$ part of the 
Hamiltonian 
as the other terms are 
constants of motion. Following same tactics as for the boson 
models we acknowledge 
the fact that the
$H_{\rm \scriptstyle XXZ}$ Hamiltonian is embedded in the $ \otimes_{j=1}^{N} 
su_{j}(2)$
algebra and make an appropriate  ansatz about the form of the
wavefunction.
Namely, we assume 
it is written in the form $\ket{z}=\otimes_{j=1}^{N} |z_j>$, where $|z_j>$ 
is the $su(2)$ coherent state correspoding to the $j$-site of the chain.  
To determine the dynamics of the complex amplitudes $z$'s we use the 
relation $z \braket{j-J_0}=\braket{J_-}$, which is 
obtained from eq.~(\ref{symb}). Then the time derivative of this relation 
for $j=1/2$, gives the evolution equation of the complex amplitude at each site 
by means
of the CS symbol of the Heisenberg equations, \viz
\bea
\dot{z}_j &=&
\frac{d/dt \braket{\sm_j} \braket{1/2 - \sth_j} - \braket{\sm_j}
d/dt \braket{1/2 - \sth_j}}
{\braket{1/2 - \sth_j}^2} \nn \\
&=& \frac{-i(\braket{[\sm_j , H]}\braket{1/2 - \sth_j} +
\braket{\sm_j}\braket{[\sth_j, H]})}
{\braket{1/2 - \sth_j}^2} \;.
\eea

\nd Straightforward evaluation of the CS symbols of the involved operators
gives the evolution equation,
\bea
i\dot{z}_j &=& \frac{1}{(1+|z_{j-1}|^2)(1+|z_{j+1}|^2)} \nn \\
& & 
\Big[
- V z_j (1-2|z_{j-1}|^2 |z_{j+1}|^2) 
+ (2g-3g |z_{j}|^2) [z_{j-1}(1+|z_{j+1}|^2)+ z_{j+1}(1+|z_{j-1}|^2)] \nn \\
&+& g z_j^2 [\overline{z}_{j-1}(1+|z_{j+1}|^2)+ 
\overline{z}_{j+1}(1+|z_{j-1}|^2)]
\Big]
\eea

\nd Returning to the original model we compute the CS mean value of 
the fermion operators by
use of eqs.~(\ref{symb}) and (\ref{jw}). This  yields
\be
\sand{z(t)}{\ad_j}{z(t)}= \prod_{k=1}^{j-1} (-2)
\sand{z_{k}(t)}{\sigma_k^{z}}{z_{k}(t)}
\sand{z_{j}(t)}{\sigma_j^{+}}{z_{j}(t)}=C_j(t)e^{-i\theta_{j}(t)}\,,
\label{fermdyn}
\ee

\nd where 
\be
C_{j}(t)=\frac{\abs{z_{j}(t)}}
{1+\absqr{z_{j}(t)}} 
\prod_{k=1}^{j-1}\frac{1-\absqr{z_{k}(t)}}{1+\absqr{z_{k}(t)}}\,,
\ee

\nd with $z_{j}(t)=\abs{z_{j}(t)}e^{i\theta_{j}(t)}$. 

\nd Also, 
$\sand{z(t)}{a_k}{z(t)}=\overline{\sand{z(t)}{\ad_k}{z(t)}}$, $k=1,\ldots ,N$,
while the average value of the fermion number operator is given by
\be
\sand{z(t)}{n_j}{z(t)}=
\sand{z(t)}{\sigma_j^z}{z(t)}+\frac{1}{2}=
\frac{\absqr{z_{j}(t)}}{1+\absqr{z_{j}(t)}}\,.
\ee

%\nd So we conclude that within the approximation
%adopted here, the mean value FP dynamics 
%is determined by the $SU(2)$-CS dynamics of the underlying XXZ model.
A final remark concerns the applicability of the CS method to the XXZ
model; it has recently come to our attention that this method also
can treat~\cite{mgm} the XXZ model of higher spins.
%In Fig.5 we plot the angles $\theta_{j}(t)$ versus
%time, which effectively gives the mean value dynamics of the fermions
%according
%to eq.~(\ref{fermdyn}). We observe that the $\theta$-phases 
%evolve nonlinearly 
%and periodically with a propagation to other neighbor sites of an 
%excitation initially localized to a single site (see figure caption).
%\begin{figure}
%\caption{Fig 5
%I'm sending you 2 more FP-plots,'figfpe' and 'figfpf'. They both
%contain the angle theta (k) (z=exp(i*theta)) versus t for a system 
%of f=5 sites with initial condition theta(1) = pi/2, theta(k) =-pi/2
%for all other k. (figfpf shows theta mod 2pi). Of course, because of
%symmetry theta(5)=theta(2) and theta(4)=theta(3) for all times.
%See my email 'mail+plots' from 20/2 concerning the difference when
%choosing f even instead of odd!}
%\bigskip
%\centerline{
%\psfig{figure=fig5.ps,height=6truecm,width=7.5truecm,angle=0}}
%\label{fig5}
%\end{figure}

\newpage
\centerline{\bf Figure captions}

Fig. 1. The Q-functions~(\ref{qfdef}) for the quintic 
$(m=3)$ GDST-trimer $(f=3)$
plotted at the time-instant $t=76.4$ for the cases of symmetric (a)
resp. normal (b) ordering of the boson operators. Dashed line
corresponds to the initially excited site $j=2$, while solid line
corresponds to the sites $j=1$ and $j=3$. (They are equal for all times
due to the symmetric initial condition.) In both cases, the total
excitation number $N=10$, the nonlinear parameter $\g=0.055$, and the
linear coupling coefficient $\l_{jk}$ is unity. 

Fig. 2. The trajectories of the centers of the Gaussian bells in
Fig. 1 for times $t < 260$. The parameter-values are the same as in
Fig. 1. Fig. 2(a,b) (Fig. 2(c,d)) corresponds to sites $1$ resp. $2$ for the SO
(NO) case.

Fig. 3. The Poissonian probability distribution $P_{n}^{j}$ 
from~eq. (\ref{poiss})
plotted as a function of excitation number $n$ and site index $j$ for the
quintic GDST $(m=3)$ with $f=21$ at the time-instant $t=50$. Fig. 3(a) shows the
case of SO, while (b) corresponds to NO. In both cases, the site $j=21$
is initially excited, and the total excitation number is $N=10$. The
value of the nonlinear parameter is $\g=0.05$, which gives
self-trapping in the SO-equation, but not in the NO-equation.

Fig. 4.
The analytic expressions of $\g_{\rm cr}$~eq. (\ref{gcr})
for a GDST
quantum dimer with quintic nonlinearity (\ie $f=2,m=3$) plotted as
a function of 
the total  excitation number $N$ for normal (NO) and symmetric (SO) ordering.
The value of the linear coupling coefficients is set equal to $1$ or to $1/2$
if we consider periodic boundary conditions. We notice that the two $\g$'s
become asymptotically equal for the moderate value of $N=10$.

%Fig. 5.
%The phase angle $\theta_j$ versus time
%for a FP system of $f=5$ sites with initial condition 
%$\theta_{1}(0) = \pi/2$, and $\theta_{j}(0) =-\pi/2$
%for $j=2,3,4,5$. Due to
%symmetry $\theta_{5}(t)=\theta_{2}(t)$ and $\theta_{4}(t)=\theta_{3}(t)$ 
%for all times. Also $J=-1$ and $\abs{z_{j}(t)}\equiv1$ for all $j$.

\newpage
\begin{thebibliography}{25}
\bibitem{bell}
J. S. Bell,  Physics (N.Y) {\bf 1},195 (1964).
\bibitem{ber}
F. A. Berezin, Comm. Math. Phys. {\bf 40}, 153 (1975).
\bibitem{xxz}
H. A. Bethe, Z. Phys. {\bf 71}, 205 (1931);
C. N. Yang and C. P. Yang, Phys. Rev. {\bf 150}, 321 (1966), and references 
therein;
M. Gaudin, Phys. Rev. Lett. {\bf 26}, 1301 (1971).
\bibitem{bbp} R. K. Bullough, N. M. Bogoliubov and G. D. Pang, in {\it Future 
Directions of Nonlinear 
Dynamics in Physical and Biological Systems}, ed. by P. L. Christiansen et al. 
(Plenum Press, New York, 1993), p. 217.
\bibitem{chjs}
P. L. Christiansen, M. H. Hays, M. F. J{\o}rgensen and A. C. Scott,
Phys. Lett. A {\bf 201}, 407 (1995), and references therein.
\bibitem{dir}
P. A. M. Dirac, {\it The Principles of Quantum Mechanics} (Oxford University 
Press,
London, 1947).
\bibitem{els}
J. C. Eilbeck, P. S. Lomdahl and A. C. Scott, Physica D {\bf 16}, 318
(1985).
\bibitem{epr}
A. Einstein , B. Podolsky and N. Rosen, Phys. Rev. {\bf 47},777 (1935)
\bibitem{de} 
D. Ellinas,
J. Phys. A: Math. Gen. {\bf 26}, L543  (1993);\\
in {\it Proc. "International Symposium on Coherent States: Past, Present
and Future"}, Eds. D. H. Feng et. al
% J. R. Klauder and
%M. R. Strayer 
(World Scientific, Singapore, 1994), p. 139;\\
in {\it Proc.  "3rd Colloquium 
on Quantum groups and Physics"}, Czech. Jour. Phys. {\bf 44},
1009 (1995).  
\bibitem{D.entang}
D. Ellinas, 
Proc. "IV Workshop on Physics and 
Computation, PhysComp96" p. 108.
Eds, T. Toffoli et. al, New England 
Complex Systems Institute 1996, also in http://pm.bu.edu/PhysComp96/.
\bibitem{ejc}
D. Ellinas, M. Johansson and P. L. Christiansen, work in progress.
\bibitem{ef}
D. Ellinas and E. G. Floratos, J. Phys. A: Math. Gen. {\bf 32} L63-L69 (1999).
\bibitem{ellman}  D. Ellinas, P. Maniadis, 
"$q$-Symmetries in DNLS-AL chains and exact solutions of quantum dimers" , quant-ph/9907014,
submitted for publication.
\bibitem{esse}
V. Z. Enol'skii, M. Salerno, A. C. Scott and J. C. Eilbeck, Physica D {\bf 59},
1 (1992).
%\bibitem{ft}
%L. D. Faddeev and L. A. Takhtajan,  {\it Hamiltonian methods in the 
%theory of solitons} (Springer-Verlag, Berlin, 1987).
\bibitem{fed}
V. K. Fedyanin and V. Yushankai, Teor. Mat. Fiz. {\bf 35}, 240 (1978);\\
V. G. Makhankov and V. K. Fedyanin, Phys. Rep. {\bf 104}, 1 (1984).
\bibitem{gil}
R. Gilmore, {\it Lie Groups, Lie Algebras and Some of Their Applications}
(J. Wiley, New York, 1974).
\bibitem{gla}
R. J. Glauber, Phys. Rev. {\bf 130}, 2529 (1963).
\bibitem{hs}
M. H. Hays and A. C. Scott, Phys. Lett. A {\bf 188}, 21 (1994).
\bibitem{h}
S. Helgason, {\it Differential Geometry, Lie Groups and Symmetric Spaces}
(Academic Press, New York, 1978).                                  
\bibitem{qfun}
T. F. Jordan and  E. C. G. Sudarshan, Rev. Mod. Phys. {\bf 33}, 515 (1962);
{\it ibid} {\bf 34}, 377 (1962);\\
N. L. Balazs and B. K. Jennings, Phys. Rep. {\bf 104} ($\#$6) 347 (1984);\\
M. Hillery, R. F. O'Connell, M. O. Scully and E. P. Wigner, Phys. Rep. 
{\bf 106} (1984).
\bibitem{kto}
G. Kalosakas, G. P. Tsironis, and E. N. Economou, J. Phys.: Condens. Mat. {\bf 
6},
7847 (1994).
\bibitem{kenkre}
V. M. Kenkre and D. K. Campbell, Phys. Rev. B {\bf 34}, 4959 (1986); 
V. M. Kenkre and G. P. Tsironis, Phys. Rev. B {\bf 35}, 1473 (1987);
G. P. Tsironis and V. M. Kenkre, Phys. Lett. A {\bf 127}, 209 (1988).
\bibitem{kib}
T. W. B. Kibble, Comm. Math. Phys. {\bf 65}, 189 (1979).
\bibitem{kla}
J. R. Klauder and B. -S. Skagerstam, {\it Coherent States} (World
Scientific, Singapore, 1985).
\bibitem{kbi}
V. E. Korepin, N. M. Bogoliubov, and A. G. Izergin, {\it Quantum
  Inverse Scattering Method and Correlation Functions} (Cambridge
University Press, 1993).
\bibitem{louis}
W. H. Louisell, {\it Quantum Statistical Properties of Radiation}, (Wiley, 
New York, 1973).
\bibitem{mgm}
V. G. Makhankov, M. Ag{\"u}ero Granados, and A. V. Makhankov,
J. Phys. A: Math. Gen. {\bf 29}, 3005 (1996).
\bibitem{per}
A. M. Perelomov,  {\it Generalized Coherent States and Their Applications} 
(Springer-Verlag, Berlin, 1985). 
\bibitem {peres} 
A. Peres, {\it Quantum Theory : Concepts and Methods }(Kluwer, Dordrecht 
1993);\\
G. Mahler  and V. A.  Weberruss, 
{\it Quantum Networks}, (Springer--Verlag Berlin 1995).  
\bibitem{pro}
J. P. Provost and G. Vallee, Comm. Math. Phys. {\bf 76}, 2890 (1980);\\
J. Anandan and Y. Aharonov, Phys. Rev. Lett. {\bf 65}, 1697 (1990).
\bibitem{pz}
F. C. Pu and B. H. Zhao, Phys. Lett. A {\bf 118}, 77 (1986).
\bibitem {risk}
H. Risken, {\it The Fokker-Planck Equation}, (Springer Berlin, 1989, 2nd ed.).
\bibitem{sch}
E. Schr\"odinger, Naturwissenschaften {\bf 14}, 664 (1926). 
\bibitem{se}
A. C. Scott and J. C. Eilbeck, Chem. Phys. Lett. {\bf 132}, 23 (1986);
Phys. Lett. A {\bf 119}, 60 (1986).
\bibitem{cs}
A. C. Scott and P. L. Christiansen, Phys. Scripta {\bf 42}, 257 (1990).
\bibitem{seg}
A. C. Scott, J. C. Eilbeck and H. Gilh{\o}j,
Physica D {\bf 78}, 194 (1994).
\bibitem{weyl}
H. Weyl, {\it The Theory of Groups and Quantum Mechanics} (Dover Publications, 
New
York, 1950).
\bibitem{wehms}
E. Wright, J. C. Eilbeck, M. H. Hays, P. D. Miller and A. C. Scott, Physica D
{\bf 69}, 18 (1993).
\bibitem{rmp}
W. M. Zhang, D. H. Feng and R. Gilmore, Rev. Mod. Phys. {\bf 62}, 867 (1990).
\end{thebibliography}
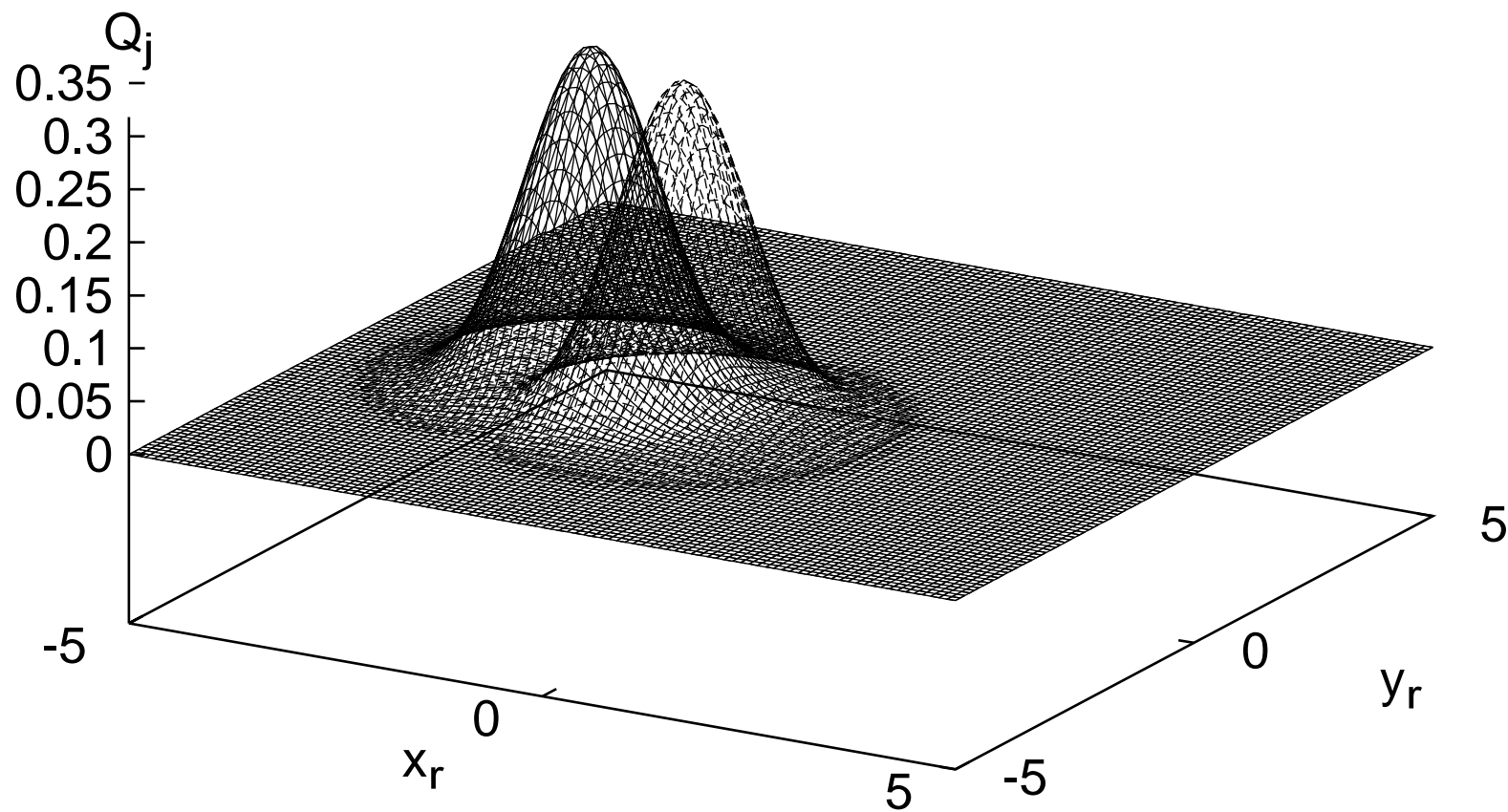
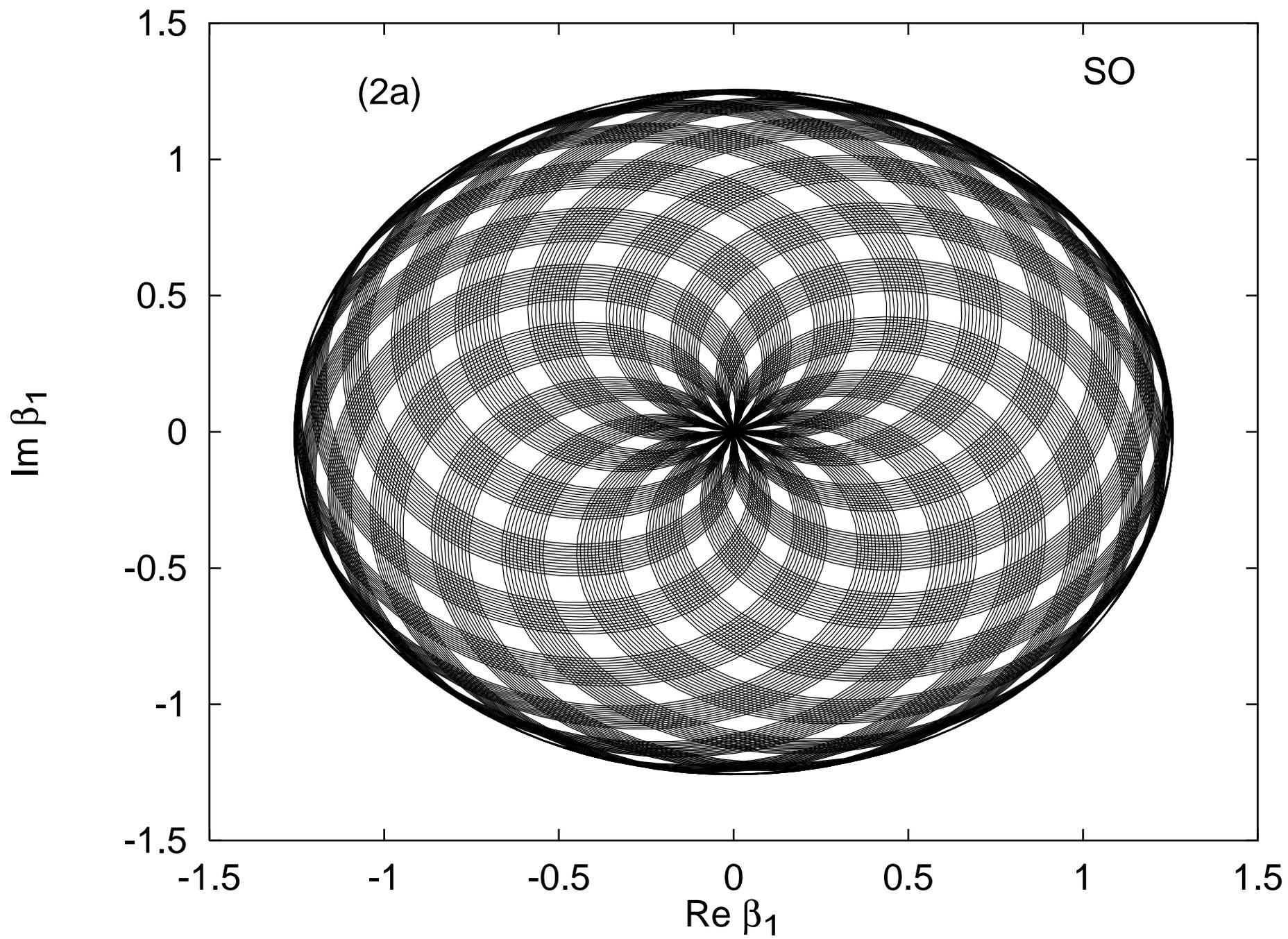
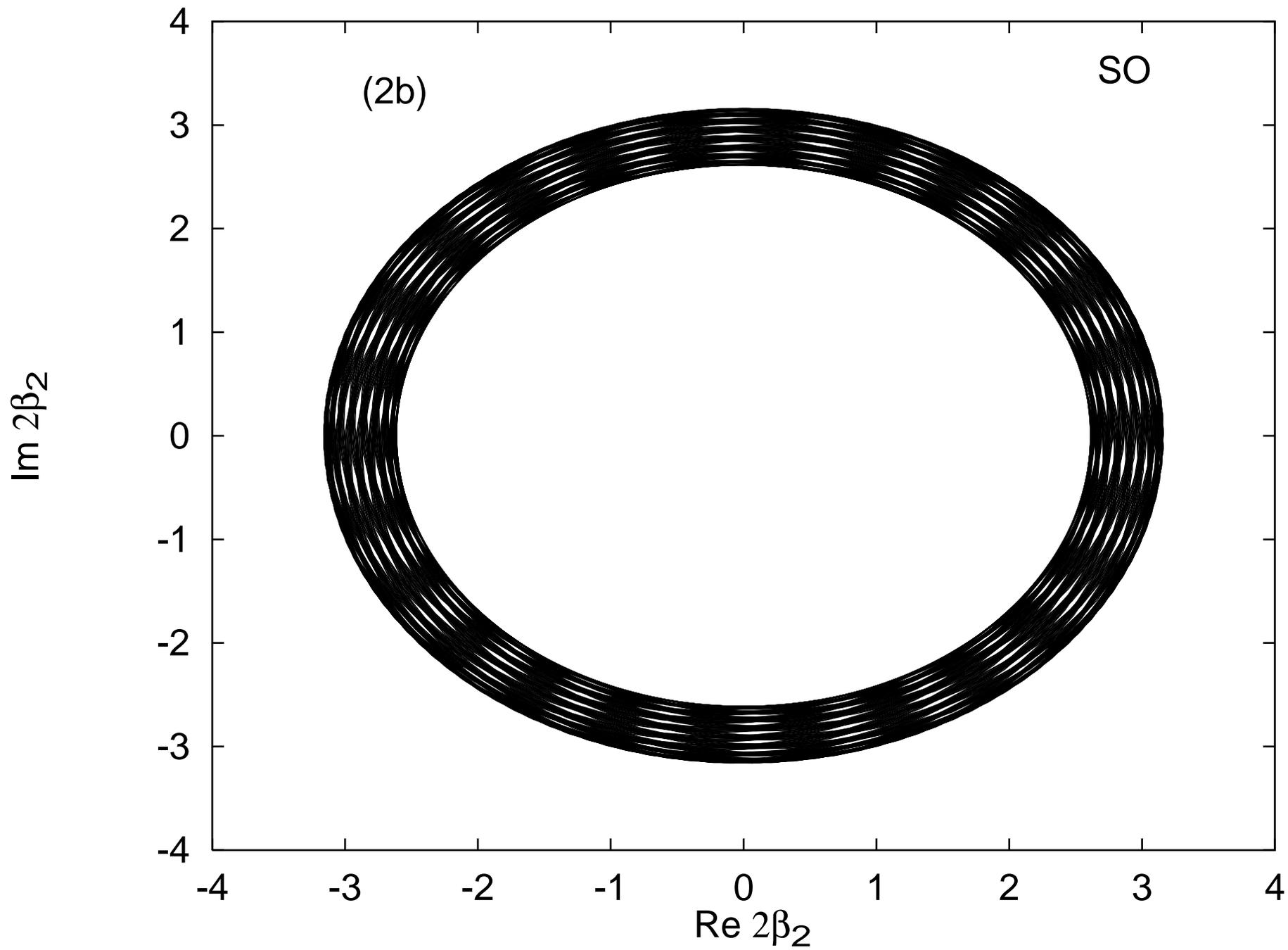
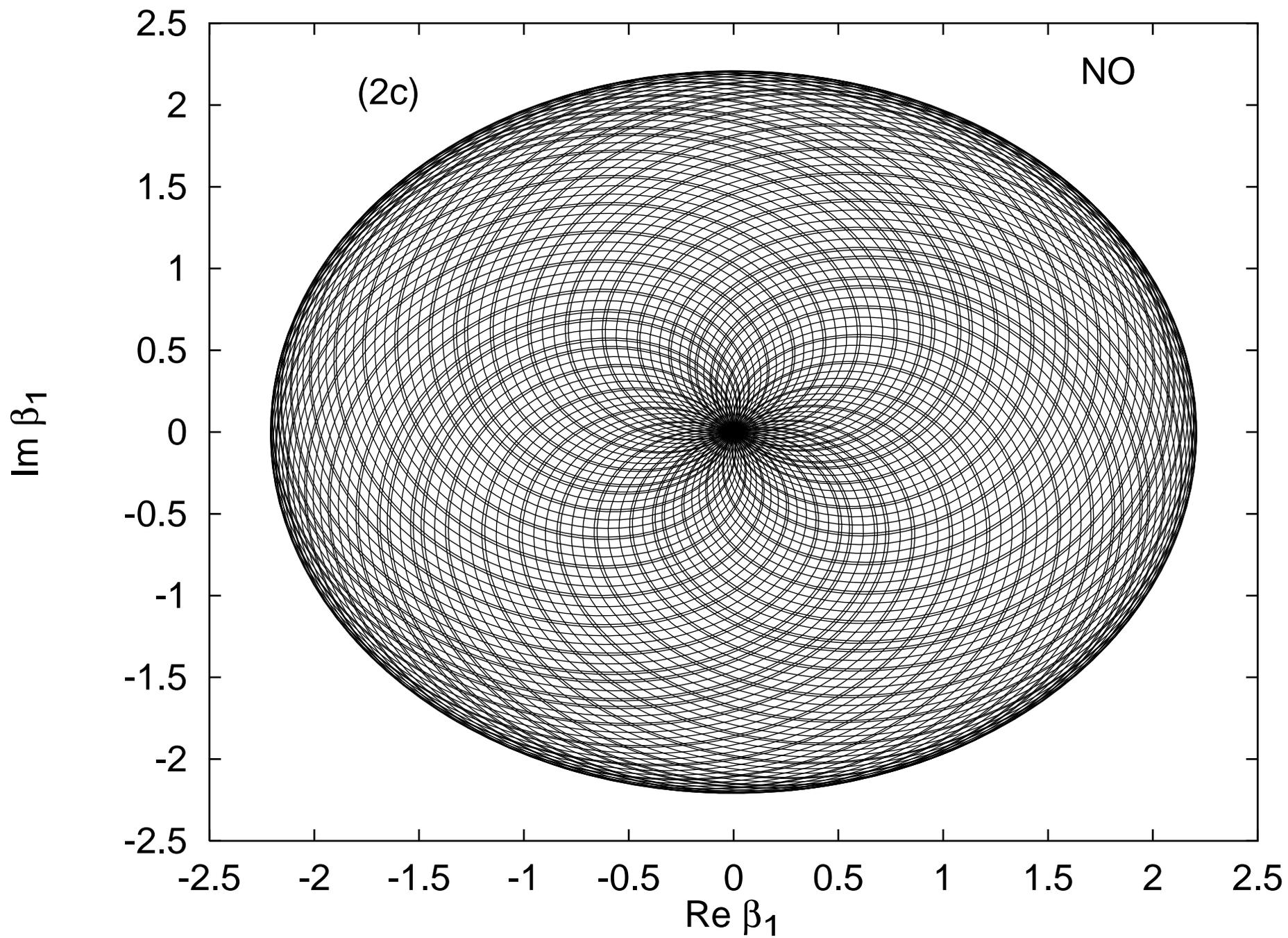
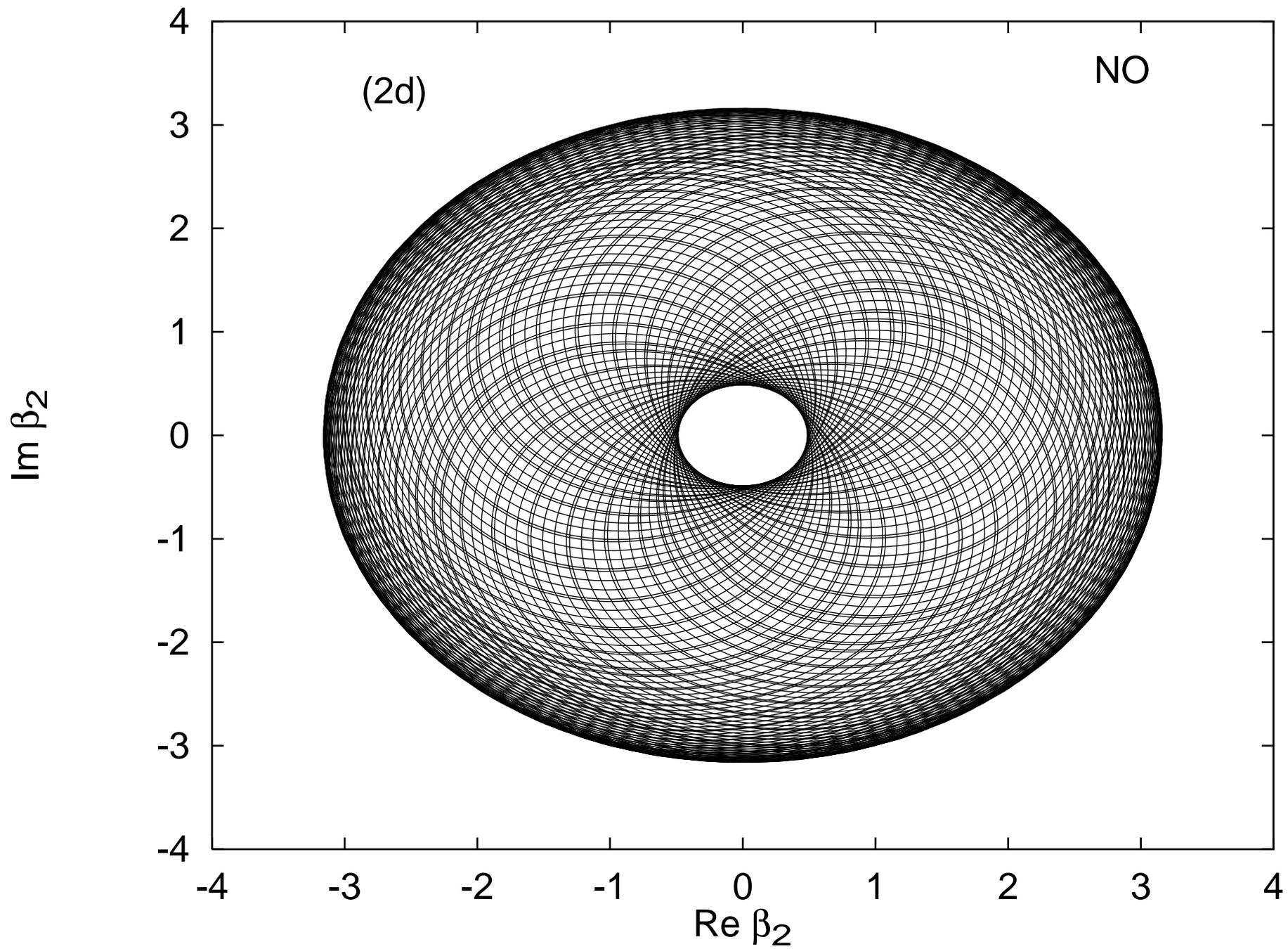
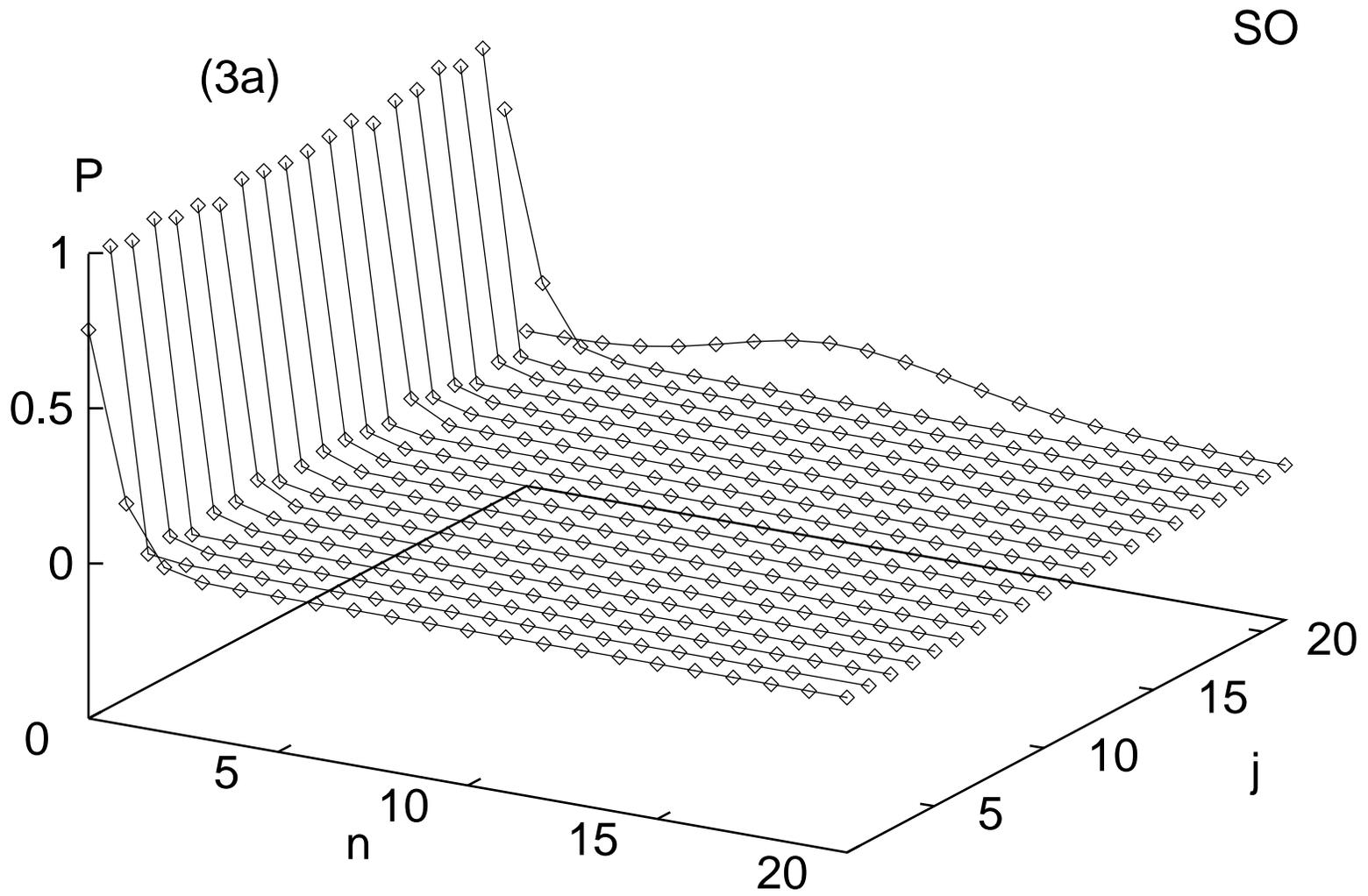
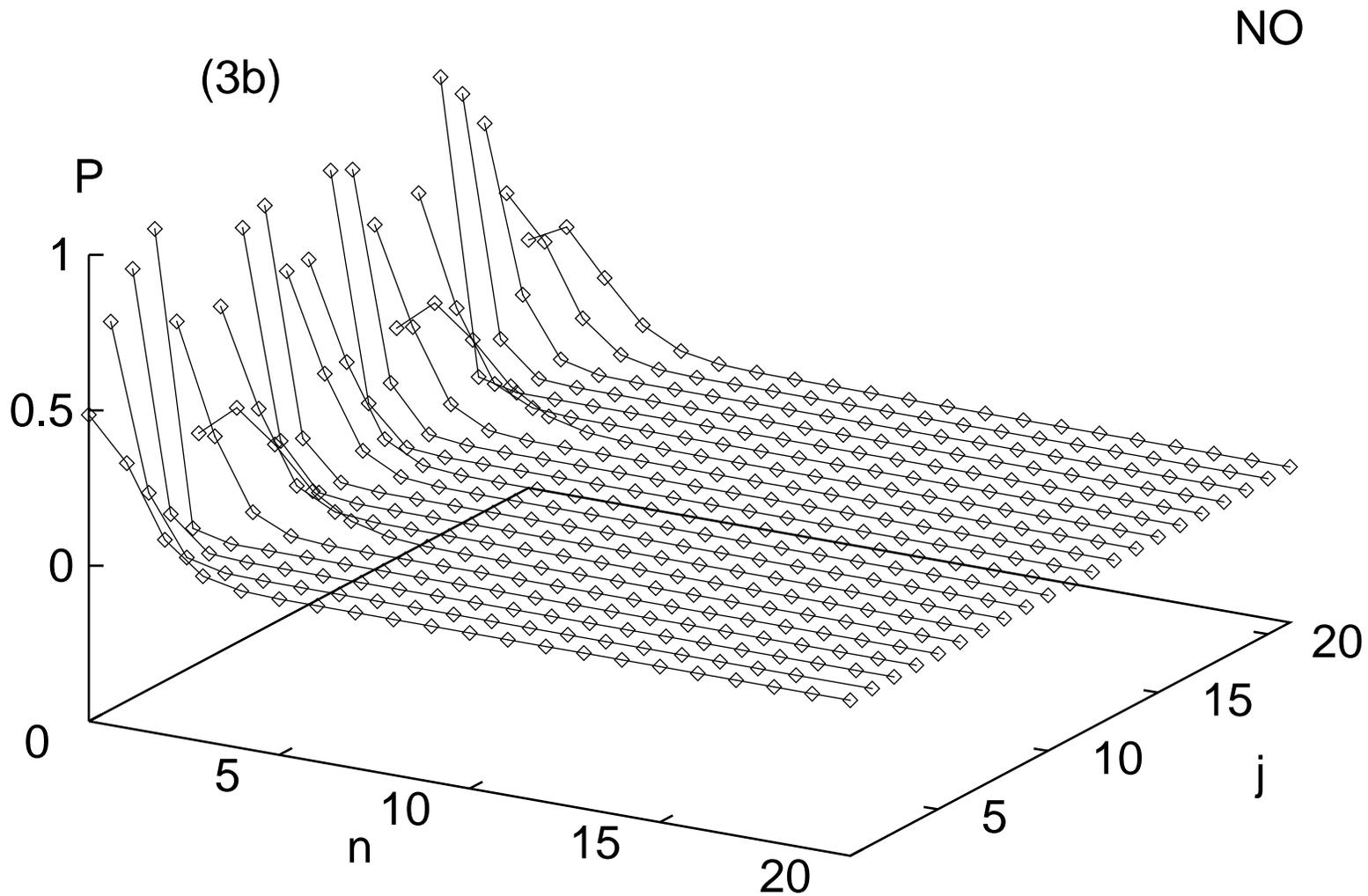
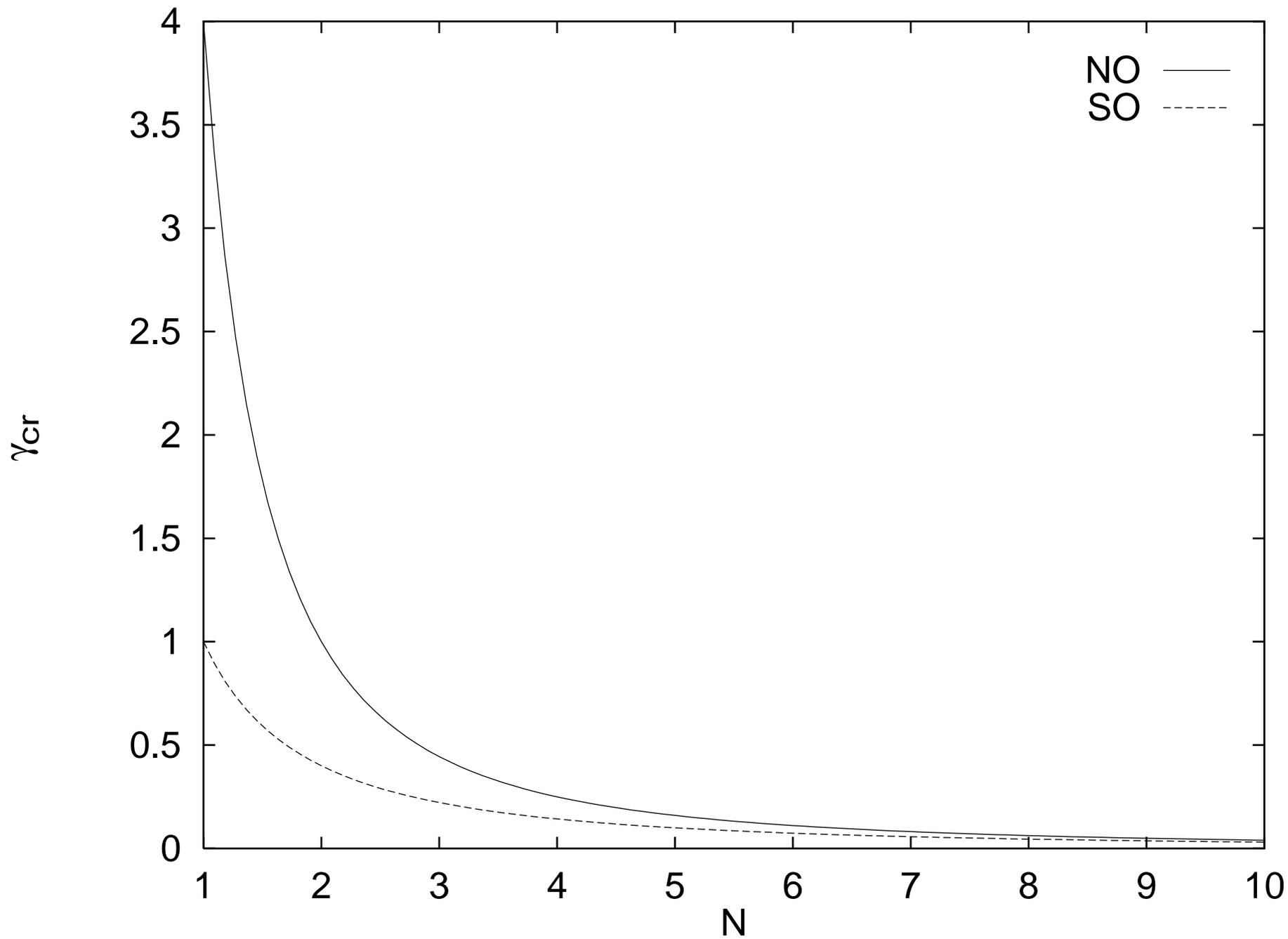
\end{document}